\documentclass[5p,number,sort&compress]{elsarticle}
\usepackage{lineno}
\usepackage{graphicx}
\usepackage[dvipdfmx,bookmarks]{hyperref}
\usepackage{gensymb}
\usepackage{units}
\usepackage{booktabs}
\usepackage{multirow}
\title{Charged particle detection performances of CMOS pixel sensors produced in a 0.18 $\mu m$ process with a high resistivity epitaxial layer}
\author[iphc]{S. Senyukov\corref{cor1}}
\author[iphc]{J. Baudot}
\author[iphc]{A. Besson}
\author[iphc]{G. Claus}
\author[iphc]{L. Cousin}
\author[iphc]{A. Dorokhov}
\author[iphc]{W. Dulinski}
\author[iphc]{M. Goffe}
\author[iphc]{C. Hu-Guo}
\author[iphc]{M. Winter}
\address[iphc]{Universit\'{e} de Strasbourg, IPHC-CNRS, 23 rue du Loess 67037 Strasbourg, France}
\cortext[cor1]{Corresponding author}
\begin{document}

\begin{abstract}
The apparatus of the ALICE experiment at CERN will be upgraded in 2017/18 during the second long shutdown of the LHC (LS2). A major motivation for this upgrade is to extend the physics reach for charmed and beauty particles down to low transverse momenta. This requires a substantial improvement of the spatial resolution and the data rate capability of the ALICE Inner Tracking System (ITS). To achieve this goal, the new ITS will be equipped with \unit[50] {$\mu m$} thin CMOS Pixel Sensors (CPS) covering either the 3 innermost layers or all the 7 layers of the detector. 
The CPS being developed for the ITS upgrade at IPHC (Strasbourg) is derived from the MIMOSA 28 sensor realised for the STAR-PXL at RHIC in a \unit[0.35] {$\mu m$} CMOS process. In order to satisfy the ITS upgrade requirements in terms of readout speed and radiation tolerance, a CMOS process with a reduced feature size and a high resistivity epitaxial layer should be exploited. In this respect, the charged particle detection performance and radiation hardness of the \emph{TowerJazz} \unit[0.18] {$\mu m$} CMOS process were studied with the help of the first prototype chip MIMOSA 32. The beam tests performed with negative pions of \unit[120]{GeV/c} at the CERN-SPS allowed to measure a signal-to-noise ratio (SNR) for the non-irradiated chip in the range between 22 and 32 depending on the pixel design. The chip irradiated with the combined dose of \unit[1]{MRad} and \unit[$10^{13}$]{$n_{eq}/cm^2$} was observed to yield a SNR ranging between 11 and 23 for coolant temperatures varying from \unit[15]{\celsius} to \unit[30]{\celsius}. These SNR values were measured to result in particle detection efficiencies above 99.5\% and 98\% before and after irradiation respectively. These satisfactory results allow to validate the \emph{TowerJazz} \unit[0.18] {$\mu m$} CMOS process for the ALICE ITS upgrade.

\end{abstract}
\maketitle
\section{Introduction}
ALICE\cite{ALJINST} is a general purpose experiment dedicated to the study of nucleus-nucleus collisions at LHC. After more than 3 years of successful operation, an upgrade of the apparatus during the second long shutdown of the LHC (LS2) in 2017/18 is under study. One of the major goals of the proposed upgrade is to extend the physics reach of the experiment for rare probes, like charmed and beauty mesons and baryons, at low transverse momenta. The efficient reconstruction of these particles requires the precise determination of the primary and secondary vertices of the collision.

Presently, it is the Inner Tracking System (ITS) composed of 6 layers of silicon sensors that is successfully used to reconstruct the vertices. However, the spatial resolution of the ITS can be significantly improved thanks to the recent developments in silicon pixel sensors and their integration. It has been demonstrated that a new ITS made of 7 layers of silicon detectors with an intrinsic point resolution of $\unit[4] {\mu m} \times \unit[4] {\mu m}$ and a material budget of $\sim$\unit[0.3] {\% $X_0$} per layer would allow to improve the impact parameter resolution by a factor of $\sim$3 \cite{ALICE_ITS_CDR}.

A dedicated and intense R\&D activity has been carried out in order to evaluate the technical feasibility of such an upgrade. As a result, CMOS pixel sensors (CPS) were chosen as the possible option to equip either 3 innermost, or all 7 layers of the new ITS. Indeed, CMOS technology allows to produce highly-granular and very light pixel sensors with a pixel pitch of about \unit[20-30]{$\mu m$} and a sensor thickness of \unit[50]{$\mu m$} satisfying the requirements of the ITS upgrade in terms of spatial resolution and material budget.

However, CPS should still prove their conformity in terms of readout speed and radiation hardness. The new ITS will be required to cope with the full interaction rate of \unit[50] {kHz} lead-lead collisions expected in LHC after the LS2, recording all minimum bias events. Such an increased event rate will lead to an important radiation load on the detector. According to recent simulations, the combined annual dose expected for the innermost layer can reach up to \unit[700]{kRad} and \unit[$10^{13}$]{$n_{eq}/cm^2$}\footnote{The expected annual dose includes also the contribution from the proton-proton collisions. These collisions will provide the reference data for the heavy-flavour analysis.}, and the operation temperature is foreseen to be $\sim$\unit[30] {\celsius}. More details on the ITS upgrade can be found in the Conceptual Design Report\cite{ALICE_ITS_CDR}.

\section{CPS developments at IPHC}
CMOS pixel sensors represent a novel but promising approach to the design of vertexing and tracking devices for particle and high-energy physics experiments. The PICSEL group of IPHC in Strasbourg has more than 12 years of experience in developing this technology. This long-term effort has recently led to the realization of the MIMOSA 28 chip to be used for the STAR-PXL detector at RHIC\cite{STAR-PXL}. MIMOSA 28 is produced with the \emph{AMS} \unit[0.35]{$\mu m$} CMOS process using a \unit[15]{$\mu m$} thick, high resistivity ($\rho > \unit[400]{\Omega \cdot cm}$) epitaxial layer. It can withstand the combined radiation dose of \unit[150]{kRad} and \unit[$3 \times 10^{12}$]{$n_{eq}/cm^2$} at the operation temperature of \unit[30] {\celsius} \cite{MIMOSA28-STAR}. The readout of the chip is performed in the rolling shutter mode with a total frame readout time of $\sim$\unit[200]{$\mu s$}.
 
A new chip being developed for the ALICE ITS upgrade has to push further the readout speed and radiation hardness limits of MIMOSA 28. Hence another CMOS process with a reduced feature size and higher resistivity epitaxial layer is desirable. The \emph{TowerJazz} \unit[0.18]{$\mu m$} CMOS process represents an attractive option providing 6 metal layers and a high-resistivity ($\rho > \unit[1]{k\Omega \cdot cm}$), \unit[18]{$\mu m$} thick, epitaxial layer. Moreover, the \emph{TowerJazz} process includes an innovative option called deep P-well. The latter represents an additional P-layer placed underneath an N-well housing a PMOS transistor. Therefore, the deep P-well guarantees in principle that the deposited signal charge is collected in the N-well containing the sensing diode and not in those housing PMOS transistors. Thus, both transistor types can be implemented in the pixels, leading to more compact in-pixel micro-circuitry.

\section{MIMOSA 32 chip}
\begin{figure}
\centering
\includegraphics[width=1\linewidth]{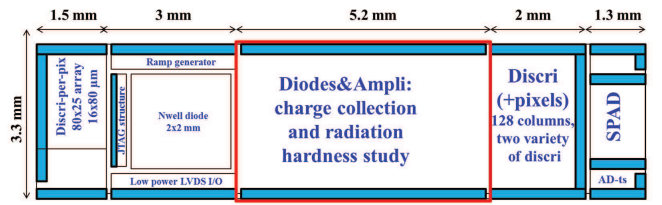}
\caption{Overview of the MIMOSA 32 chip components.}
\label{fig:MI32}
\end{figure}

MIMOSA 32 is the first exploratory prototype chip designed at IPHC, in collaboration with IRFU (Saclay), investigating the charged particle detection performances of the \emph{TowerJazz} \unit[0.18] {$\mu m$} CMOS process for subatomic physics. The chip was submitted for production in December 2011. It returned back from the foundry in March 2012. As presented on \autoref{fig:MI32}, MIMOSA-32 consists of several independent blocks aimed for the study of different components of a pixel sensor, like charge collection, pre-amplification, binary encoding and steering circuits. The study presented in this paper was performed on the central part of the chip referred to as "Diodes \& Ampli" in \autoref{fig:MI32}, covering a surface of \unit[$5.2\times 3.3$]{$ mm^2$}. This part contains 32 submatrices featuring different pixel architectures and sizes. 22 submatrices contain simply a sensing diode connected to a source follower, while the 10 other submatrices additionally include a pre-amplification chain. 28 submatrices have square pixels of \unit[$20\times 20$]{$\mu m^2$}. Two other pairs of submatrices are composed of rectangular pixels of \unit[$20\times 40$]{$\mu m^2$} and \unit[$20\times 80$]{$\mu m^2$} respectively. The readout chain allows to read out one submatrix at a time in the rolling shutter mode. The total submatrix readout time is \unit[32]{$\mu s$} at a clock frequency of \unit[2]{MHz}.

\section{Irradiation and tests}
Upon arrival from the foundry, a sample of the chips was irradiated with 10 keV X-Rays at CERN and with the $\sim$\unit[1]{MeV} neutrons delivered by the FRM II reactor near Munich. The applied ionizing and non-ionizing loads ranged from \unit[300]{kRad} to \unit[10]{MRad} and from \unit[$3 \times 10^{12}$]{$n_{eq}/cm^2$} to \unit[$3 \times 10^{13}$]{$n_{eq}/cm^2$} respectively. Several chips received a combined dose of \unit[1]{MRad} and \unit[$10^{13}$]{$n_{eq}/cm^2$} representing approximately the expected dose for the upgraded ALICE ITS after several years of running.

All non-irradiated and irradiated chips were first tested in the laboratory at IPHC with a $Fe^{55}$ source. All the tests were performed at two different temperatures of the chip: \unit[15]{\celsius} and \unit[30]{\celsius}. These measurements allowed to determine the pixel noise of different submatrices and their charge collection efficiency (CCE) as a function of the irradiation dose and chip temperature. On the basis of these results the best performing submatrices were selected for beam tests.

The beam tests of MIMOSA 32 were performed at the CERN-SPS during Summer 2012. A micro-strip telescope\cite{Strip_telescope} composed of 8 reference planes was installed on the T4-H6 beam line delivering negative pions with an energy of $\sim$\unit[120]{GeV/c}. Triggering was done by the coincidence signal of two scintillators with an overlapping area of $\unit[2\times 2]{mm^2}$. The chip being tested (Device Under Test or DUT) was connected to the cooling system, giving possibility to control the sensor temperature during the tests.
 
\section{Analysis}
Data taken during the beam test was used to measure two essential performance parameters: the signal-to-noise ratio (SNR) and the particle detection efficiency.

The analysis algorithm was the following. First, the SNR was calculated for all the pixels in the DUT. The pixels with the highest SNR were used as seeds for the clusters reconstruction. Next, a cut was applied accepting only clusters with a seed pixel SNR higher than 5. At the next step an association was sought between selected clusters and reference telescope tracks. The cluster was considered as associated to the track if its centre of gravity was located closer than \unit[100]{$\mu m$} to the extrapolated track position in the DUT plane. The distribution of the seed pixel SNR of all the associated clusters was built. This distribution was then fitted with the Landau's function and characterized by its most probable value (MPV).

The particle detection efficiency was calculated as the ratio between the number of reference telescope tracks passing through the DUT and the number of good clusters associated to these tracks.
\section{Beam test results}
\begin{figure}
\includegraphics[width=\linewidth]{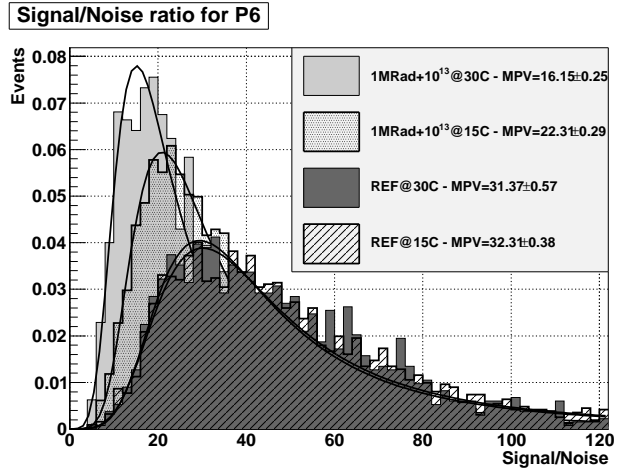}
\caption[]{Signal-to-noise ratio distributions for cluster seed pixels composing the submatrix with simple square pixels of \unit[$20 \times 20$] {$\mu m^2$} (P6) at \unit[15]{\celsius} and \unit[30]{\celsius}, before and after irradiation with the combined dose of \unit[1] {MRad} and \unit[$10^{13}$] {$n_{eq}/cm^2$}.
}
\label{fig:P6_SNR}
\end{figure}

\begin{figure}
\includegraphics[width=\linewidth]{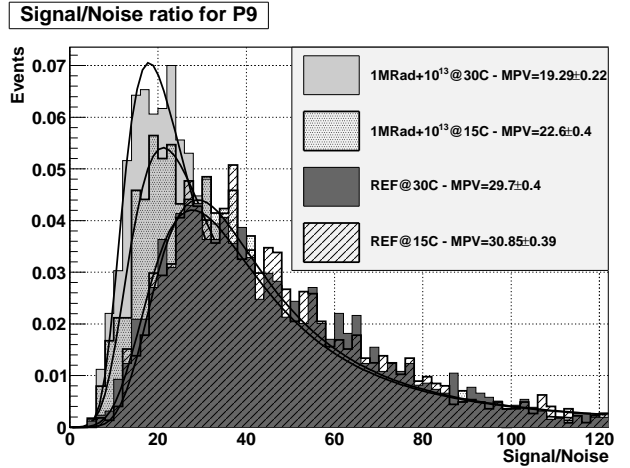}
\caption[]{Signal-to-noise ratio distributions for cluster seed pixels composing the submatrix with square pixels of \unit[$20 \times 20$] {$\mu m^2$} including a deep P-well (P9) at \unit[15]{\celsius} and \unit[30]{\celsius}, before and after irradiation with the combined dose of \unit[1] {MRad} and \unit[$10^{13}$] {$n_{eq}/cm^2$}. 
}
\label{fig:P9_SNR}
\end{figure}

\begin{figure}
\includegraphics[width=\linewidth]{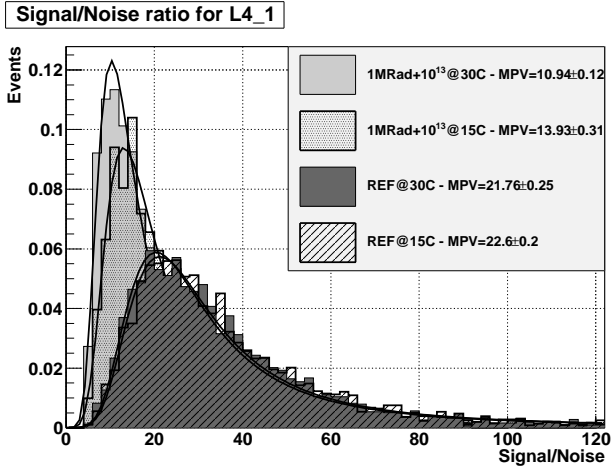}
\caption[]{Signal-to-noise ratio distributions for cluster seed pixels composing the submatrix with simple rectangular pixels of \unit[$20 \times 40$] {$\mu m^2$} (L41) at \unit[15]{\celsius} and \unit[30]{\celsius}, before and after irradiation with the combined dose of \unit[1] {MRad} and \unit[$10^{13}$] {$n_{eq}/cm^2$}.
}
\label{fig:L41_SNR}
\end{figure}

In the following, results for the three submatrices composed of the pixel types most relevant for the ALICE ITS upgrade are presented. They first address the question of the charge collection efficiency and noise performance with a relatively small pixel (\textbf{P6}), which should be optimal in terms of detection efficiency because of its relatively large sensing diode surface and dense diode implantation. The next question considered concerns the potential perturbation induced by the deep P-wells (\textbf{P9}). Finally, rectangular pixels (\textbf{L41}) were used to investigate the possibility of elongating the pixels in order to reduce the readout time, focusing on the potential detection efficiency deterioration due to the reduced sensing node density. 
\begin{description}
\item[P6] -- simple square pixels of \unit[$20 \times 20$]{$\mu m^2$} with a diode surface of \unit[10.9]{$\mu m^2$}. This simplest structure was used as the reference. SNR distributions for the seed pixel and corresponding MPV values for the submatrix P6, at \unit[15]{\celsius} and \unit[30]{\celsius}, before and after irradiation with the combined dose of \unit[1] {MRad} and \unit[$10^{13}$] {$n_{eq}/cm^2$}, are shown in \autoref{fig:P6_SNR}.

\item[P9] -- square pixel of \unit[$20 \times 20$]{$\mu m^2$} with a diode surface of \unit[10.9]{$\mu m^2$} featuring a deep P-well. SNR distributions of the seed pixel and corresponding MPV values for the submatrix P9, at \unit[15]{\celsius} and \unit[30]{\celsius}, before and after irradiation with the combined dose of \unit[1] {MRad} and \unit[$10^{13}$] {$n_{eq}/cm^2$}, are displayed in \autoref{fig:P9_SNR}.

\item[L41] -- simple rectangular pixel of \unit[$20 \times 40$]{$\mu m^2$} with a diode surface of \unit[9]{$\mu m^2$}. SNR distributions of the seed pixel and corresponding MPV values for the submatrix L41, at \unit[15]{\celsius} and \unit[30]{\celsius}, before and after irradiation with the combined dose of \unit[1] {MRad} and \unit[$10^{13}$] {$n_{eq}/cm^2$}, are presented in \autoref{fig:L41_SNR}.
\end{description}

All three sets of the SNR distributions share some similar features. Firstly, the non-irradiated chip is virtually insensitive to the temperature in the range considered, as the corresponding SNR distributions (dark grey and hatched histograms) almost completely overlap. Secondly, the irradiated chip became sensitive to the temperature variations. Indeed, the SNR distributions at \unit[30]{\celsius} (light grey histograms) are all shifted towards lower values with respect to those at \unit[15]{\celsius} (dotted histograms). A separate measurement of the charge and the noise distributions showed that the SNR decrease induced by the radiation load is mainly due to the non-ionizing component and the temperature dependence originates essentially from the bulk leakage current.

Another important feature is the nearly equivalent SNR for the \textbf{P6} and \textbf{P9} submatrices, indicating that the deep P-well does not introduce parasitic charge collection, even after irradiation.

Finally, the rectangular pixels (\textbf{L41}) exhibit significantly lower SNR values than the square pixels (\textbf{P6}, \textbf{P9}). It can be explained by two main factors. First, the distance between neighbouring collecting diodes is $ \sqrt{2} $ times bigger than that of square pixels. Secondly, the diode of the \textbf{L41} submatrix has a surface of only \unit[9]{$ \mu m^2 $}, while it amounts to \unit[10.9]{$ \mu m^2 $} for \textbf{P6} and \textbf{P9}. These two factors lead to less efficient charge collection by the rectangular pixels and therefore lower SNR\footnote{The charge collection efficiency was measured with a $Fe^{55}$ source for all three pixel designs. It was found to reflect the SNR differences observed between \textbf{L41} and \textbf{P6} and \textbf{P9}.}.

However, the particle detection efficiency has to be additionally taken into account in order to draw the conclusions on the performance. Particle detection efficiency for the three pixel types in different conditions, together with the corresponding SNR values, are reported in \autoref{tab:summary}. Several observations can be made. The square pixel structures provide a good detection efficiency, higher than \unit[99.5]{\%} at both temperature regimes, before and after irradiation. Meanwhile the rectangular pixels have shown a different behaviour. On the one hand, before irradiation, submatrix \textbf{L41} provides an excellent detection efficiency of about \unit[99.8]{\%} at both temperatures despite the lower SNR values. On the other hand, after irradiation, the detection efficiency drops to $\sim$\unit[98]{\%} at \unit[30]{\celsius} while not changing significantly at \unit[15]{\celsius}. Though moderate, this efficiency loss could be mitigated with an optimization of the charge sensing system design and with an adequate control of the operating temperature.

\begin{table*}[ht]
\caption{Signal-to-noise ratio (SNR) and particle detection efficiency for the three pixels types at \unit[15]{\celsius} and \unit[30]{\celsius}, before and after irradiation. Uncertainties are statistical only.}
\label{tab:summary}
\centering
\begin{tabular}{ccccc}
\toprule
 Pixel type & Radiation load &  Temperature & SNR &  Efficiency [\%] \\
 \midrule
 \multirow{4}{*}{P6} & \multirow{2}{*}{---}     & 15 \celsius& $32.31\pm 0.38$ & $99.84\pm 0.07$\\ 
                     &                                                      & 30 \celsius & $31.37\pm 0.57$ & $99.64\pm 0.16$\\ 
 \cmidrule(rl){2-5} 
                     & \multirow{2}{*}{1 MRad + $10^{13}$ $n_{eq}/cm^2$}    & 15 \celsius & $22.31\pm 0.29$ & $99.87\pm 0.08$\\ 
                     &                                                      & 30 \celsius & $16.15\pm 0.25$ & $99.77\pm 0.11$\\ 
 \midrule
 \multirow{4}{*}{P9} & \multirow{2}{*}{---}& 15 \celsius & $30.85\pm 0.39$ & $99.91\pm 0.06$\\ 
                     &                                                      & 30 \celsius & $29.70\pm 0.40$   & $99.74\pm 0.10$\\ 
 \cmidrule(rl){2-5} 
                     & \multirow{2}{*}{1 MRad + $10^{13}$ $n_{eq}/cm^2$}    & 15 \celsius & $22.60\pm 0.40$   & $99.92\pm 0.08$\\ 
                     &                                                      & 30 \celsius & $19.29\pm 0.22$ & $99.87\pm 0.07$\\
 \midrule
 \multirow{4}{*}{L41}& \multirow{2}{*}{---}      & 15 \celsius & $22.60\pm 0.20$   & $99.86\pm 0.06$\\ 
                     &                                                      & 30 \celsius & $21.76\pm 0.25$ & $99.78\pm 0.08$\\ 
 \cmidrule(rl){2-5}  
                     & \multirow{2}{*}{1 MRad + $10^{13}$ $n_{eq}/cm^2$}    & 15 \celsius & $13.93\pm 0.31$ & $99.51\pm 0.25$\\ 
                     &                                                      & 30 \celsius & $10.94\pm 0.12$ & $97.99\pm 0.25$\\ 
\bottomrule
\end{tabular}
\end{table*}

\section{Conclusions}
Preliminary results obtained with the MIMOSA 32 chip demonstrated a good potential of the \emph{TowerJazz} 0.18 $\mu m$ CMOS process for charged particle detection. The signal-to-noise ratio of the non-irradiated sensor reached values up to $ 32.31 \pm 0.38 $ at \unit[15]{\celsius}. The corresponding particle detection efficiency is $ 99.84\pm 0.07 $ \%. Pixel structures featuring a deep P-well showed no significant degradation of the charge collection efficiency independently of the radiation load. This will allow exploiting this process option in the design of the upcoming sensors.

Finally, the performance obtained with the rectangular pixels shows that they can be used for the ALICE ITS upgrade program. The chip irradiated with the combined dose of \unit[1]{MRad} and \unit[$10^{13}$] {$n_{eq}/cm^2$}, which correspond to several years of operation, provides a detection efficiency which is still equal to $97.99\pm 0.25$ \% at a temperature of \unit[30]{\celsius}. Accounting for the possibility of further optimizations of the chip circuitry and of an operating temperature below \unit[30]{\celsius}, the ultimate detection efficiency is likely to reach at least 99 \%.

The successful validation of the charge collection performance of the \emph{TowerJazz} \unit[0.18] {$\mu m$} CMOS process will soon be followed by the development and study of the other sensor building blocks, like in-pixel pre-amplification, signal discrimination and sparse data scan produced in this process.

The sensor being developed for the ALICE ITS targets specifications (e.g. \unit[15-30]{$\mu s$} readout time) which are relevant for other experiments, like CBM-MVD, ILD-VXD, etc. The first results obtained with the \emph{TowerJazz} \unit[0.18] {$\mu m$} CMOS process addressed in this paper indicate that it is well suited to their requirements.

\bibliographystyle{elsarticle-num}
\bibliography{Senyukov-RESMDD12}
\end{document}